\documentclass[12pt]{article}
\usepackage[cp866]{inputenc}

\usepackage{amsfonts,amsmath,amssymb}
\usepackage{amssymb}
\usepackage{epsfig}

\begin{document}\def\p{\phi}\def\P{\Phi}\def\a{\alpha}\def\e{\epsilon}
\def\be{\begin{equation}}\def\ee{\end{equation}}\def\l{\label}
\def\0{\setcounter{equation}{0}}\def\b{\beta}\def\S{\Sigma}\def\C{\cite}
\def\r{\ref}\def\ba{\begin{eqnarray}}\def\ea{\end{eqnarray}}
\def\n{\nonumber}\def\R{\rho}\def\X{\Xi}\def\x{\xi}\def\la{\lambda}
\def\d{\delta}\def\s{\sigma}\def\f{\frac}\def\D{\Delta}\def\pa{\partial}
\def\Th{\Theta}\def\o{\omega}\def\O{\Omega}\def\th{\theta}\def\ga{\gamma}
\def\Ga{\Gamma}\def\t{\times}\def\h{\hat}\def\rar{\rightarrow}
\def\vp{\varphi}\def\inf{\infty}\def\le{\left}\def\ri{\right}
\def\foot{\footnote}\def\vep{\varepsilon}\def\N{\bar{n}(s)}
\def\k{\kappa}\def\sq{\sqrt{s}}\def\bx{{\mathbf x}}\def\La{\Lambda}
\def\bb{{\bf b}}\def\bq{{\bf q}}\def\cp{{\cal P}}\def\tg{\tilde{g}}
\def\cf{{\cal F}}\def\bN{{\bf N}}\def\Re{{\rm Re}}\def\Im{{\rm Im}}
\def\bk{\mathbf{k}}\def\cl{{\cal L}}\def\cs{{\cal S}}\def\cn{{\cal N}}
\def\cg{{\cal G}}\def\q{\eta}\def\ct{{\cal T}}\def\bbs{\mathbb{S}}
\def\bU{{\mathbf U}}\def\bE{{\mathbf e}}\def\bc{{\mathbf C}}
\def\vs{\varsigma}\def\cg{{\cal G}}\def\ch{{\cal H}}\def\df{\d/\d }
\def\mz{\mathbb{Z}}\def\ms{\mathbb{S}}\def\kb{{\mathbb
K}}\def\cd{\mathcal D}\def\mj{\mathbf{J}}\def\Tr{{\rm Tr}}
\def\bu{{\mathbf u}}\def\by{{\mathrm y}}\def\bp{{\mathbf p}}
\def\k{\kappa} \def\cz{{\mathcal Z}}\def\ma{\mathbf{A}}
\def\me{\mathbf{E}}\def\mp{\mathbf{P}}\def\ra{\mathrm{A}}
\def\ru{\mathrm{u}}\def\rP{\mathrm{P}}\def\rp{\mathrm{p}}\def\z{\zeta}
\def\my{\mathbf Y}

\begin{center}
{\huge\bf On elimination of the Gribov ambiguity}

\vskip 0.5cm {\large\it J.Manjavidze\foot{Inst. Phys. (Tbilisi,
Georgia) $\&$ JINR (Dubna, Russia), E-mail: joseph@nu.jinr.ru},
A.Sissakian\foot{JINR (Dubna, Russia), E-mail: sisakian@jinr.ru} and
V.Voronyuk\foot{JINR (Dubna, Russia), E-mail: vadimv@nu.jinr.ru}}

\end{center}

\begin{abstract}
We find a strong coupling expansion around the non-trivial
extremum of the Yang-Mills action. It is shown that the developed
formalism is the Gribov ambiguity free since each order of the
developed perturbation theory is transparently gauge invariant.
The success is a consequence of the restriction: calculations are
not going beyond the norm of the $S$-matrix element.

\end{abstract}
%\newpage
\tableofcontents%
\newpage

\section{Introduction}\0

In the late seventieth V.Gribov discovered  that it was impossible
to extract unambiguously the non-Abelian gauge symmetry degrees of
freedom if the gauge field was strong with the Coulomb gauge to be
applied \C{gribov}. It was shown later that the same phenomenon
appeared in the arbitrary gauge fixing conditions \C{atiyah,
zinger}. On the other hand, the canonical quantization scheme
certainly prescribes to extract the symmetry degrees of freedom
\C{dirak}.

This problem is unavoidable, Sec.2.2, in the general approach to
the Yang-Mills theory, but it can be resolved calculating the norm
of the amplitude, $|\mz|$, Sec.4.4.

Formally, we present a partial solution of the problem since the
phase of $\mz$ is excluded from consideration. Then the result of
Gribov, Atiyah and Singer \C{gribov, atiyah, zinger} might signify
that the problem is unsolvable in the presence of the phase. In
other words, we will argue that the Yang-Mills field theory is
free from ambiguities if it is used for the description of
observables: in this sense it can exist and that will do. The
application of this formalism restricted by the norm was deduced
in a number of papers, see e.g. \C{elpat, physrep}.

\subsection{\it Solution of the Gribov problem in brief}

It is necessary to give a more exact answer to the question: what
does the arbitrariness of the phase mean? The optical theorem: \be
i\mz\mz^\dag=(\mz-\mz^\dag),\l{1.1a}\ee  will be involved for this
purpose \C{jmp-1}. Therefore, the wish to leave the phase
arbitrary means that only the absorption part of amplitudes,
$\D\mz=(\mz-\mz^\dag)/2i$, would be the object of our
calculations.

The functional integral representation of $\D\mz$ is unknown. For
this reason we will start with $\mz$ and find the functional
integral representation for $\D\mz$ using (\r{1.1a}).

It was shown that $\D\mz$ is defined on the $\d$-like (Dirac)
functional measure \C{yaph}: \be DM(A)=\prod_{x}dA_{a\mu}(x)\d\le(
\f{\d S (A)}{\d A^{a\mu}(x)}+\hbar J_{a\mu}(x)\ri),\l{1.1}\ee
where $A_{a\mu}$ is the Yang-Mills potential, $a$ is the colour
index. The reason of Dirac measures appearance is the cancellation
of contributions in the difference of r.h.s. of (\r{1.1a}). They
are "unnecessary" from the point of view of conservation of total
probability. The boundary condition that the end points of action
do not vary, (\r{2.16a}), has been used. Sec.2. contains the
derivation of (\r{1.1}) for the Yang-Mills theory.

We have found that the corresponding quantum perturbations are
excited by the Gauss operator $\exp\{-i\kb(J)\}$ in the vicinity
of $J_{a\mu}=0$, see Eq.(\r{2.22a}). It can be shown \C{jmp-1}
that the theory restores in full measure the "weak-coupling"
expansion of the type described in the papers \C{jackiw-1,
jackiw-2, faddeev}.

The "correspondence principle" written in (\r{1.1}) is strict, it
does not depend on the value of Plank constant $\hbar$. Thus, it
defines the rule how the quantum force, $J_{a\mu}(x)$, must be
transformed under the transformation of $A_{a\mu}(x)$. The latter
is impossible in the general functional integral representation of
$\mz$ \C{edwards, marinov}, see also \C{shabanov}.

The Dirac measure orders to perform the transformation in the
class of strict solutions, $u_{a\mu}(x)$, of the sourceless (with
$J_{a\mu}=0$) Lagrange equation. This stands for \C{hioe, isham,
linden, landsman, mackey} mapping into the coset space $\cg/\ch$:
\be u_{a\mu}:~~A_{a\mu}(x)\to \{\la\}\in \cg/\ch,\l{1.2}\ee where
$\cg$ is the symmetry group of the problem and $\ch$ is the
invariance group of $u_{a\mu}$. The qualitative reason of this
choice is the following: after having got the ground state field,
$\forall u_{a\mu}(x)$, the freedom in the choice of the value of
integration constants, $\{\la\}$, is what remains from the
continuum of the field degrees of freedom. The gauge phases
$\{\La^a\}\subset\{\la\}$.

The mapping (\r{1.2}) involves the splitting \C{jmp-1}: \be
J_{a\mu}(\bx,t)\to (j_\x,j_\q)(t),~\{\x,\q\}\in T^*W,\l{1.3}\ee
where\foot{ The developed formalism will not be manifestly Lorentz
covariant. The space component $\bx$ and time $t$,
$x^2=t^2-\bx^2$, were decoupled for this reason.} the symplectic
subspace $T^*W\subseteq W$ and $W$ is the physical coset space.
The definition of the physical coset space is given in Sec.2.2.

The transformed perturbations generating operator,
$\exp\{-i\bk(j)\}$, is still Gaussian. It acts in $T^*W$ and
generates the "strong-coupling" perturbation series if
$u_{a\mu}\sim1/\sqrt g$, $g$ is the interaction constant. The
question of the existence of perturbation series of such a type
presents a separate problem. We will assume that the series exist.
The expansion of the operator exponent $\exp\{-i\bk(j)\}$ produces
also the asymptotic series over the non-negative even powers of
$\hbar$ \C{tmf}.

It will be shown that $\{\La^a(\bx,t)\}\nsubseteq \{\x,\q\}$.
Therefore, each order of the expansion of $\exp\{-i\bk(j)\}$ is
the transparently gauge invariant quantity and no gauge fixing
procedure will be required. This is the main result. The
preliminary verse of it was given in \C{jmp-3}.

\subsection{\it Agenda}

We will draw attention to the following two questions.

We will find that \be W=T^*W+R,\l{1.4}\ee where $\{\La^a\}\in R$
since there is no conjugate to $\La^a$ gauge charge dependence in
the field $u_{a\mu}$.

The mapping (\r{1.2}) is singular since $\dim T^*W<\infty$. We will
show that the singularity can be isolated and cancelled by the
normalization. This result allows to conclude that no gauge fixing
problem will arise because of the renormalization procedure.

The structure of the paper is given in the table of Contents.

\section{Perturbation theory}

We will consider the theory with the action:\be S(A)=-\f{1}{4}\int
d^4x F_{\mu\nu}^a(A)F^{\mu\nu}_a(A) \l{f3}\ee The Yang-Mills
fields \be F_{a\mu\nu}(A)=\pa_\mu A_{a\nu}-\pa_\nu A_{a\mu}-g
C_a^{bc}A_{b\mu}A_{c\nu} \l{f4}\ee are the covariants of the
non-Abelian gauge transformations. The group will not be
specified. The matrix notation: $A_{a\mu}\o_a=A_{\mu}$ will be
also used.

We will calculate the quantity\foot{The generalization was
considered in \C{physrep, tmf}.}: \be \cn=|\cz|^2,\l{}\ee where
\be \cz=\int DA e^{iS_C(A)},~DA=\prod_{\bx,t\in
C}\prod_{a,\mu}\f{d A_{a,\mu}(\bx,t)}{\sqrt{2\pi}},\l{}\ee is
defined on the Minkowski metric. The Mills complex time formalism
will be used to avoid the light cone singularities \C{milss}. For
example, \be C:~t\to t+i\vep,~\vep\to+0,~-\infty\leq
t\leq+\infty.\l{}\ee At the very end one must take $\vep=+0$. The
Mills formalism restores the Feynman's $i\e$-prescription.

\subsection{\it Dirac measure}

The double integral: \be \cn=\int DA^+DA^-e^{iS_C(A^+)-
iS^*_C(A^-)}\l{2.12}\ee will be calculated using the equality
(\r{1.1a}). To extract the Dirac measure\foot{The term "$\d$-like
(Dirac) measure" have been taken from \C{fedor}.}, one must
introduce the mean trajectory, $A_{a\mu}$, and the virtual
deviation, $a_{a\mu}$, instead of $A_\mu^\pm$: \be A_{a\mu}
^\pm(x)=A_{a\mu}(x)\pm a_{a\mu}(x). \l{2.13}\ee It will be shown
that the matrix $a_{\mu}$ is the covariant of gauge
transformations: \be a_\mu\to a_\mu'=\O a_\mu\O^{-1}\l{2.14}\ee
The transformation (2.12) is linear and the differential measure
\be DA^+DA^-=\prod_{\bx,t\in C+C^* }\prod_{a,\mu}dA_{a\mu}(x)
\prod_{\bx,t\in C+C^*}\prod_{a,\mu}\f{da_{a\mu}(x)} {\pi} \equiv
DADa.\l{2.15a}\ee is defined on the complete time contour $C+C^*$.
Notice that \be A_{a\mu}(\bx,t\in C^*)=A_{a\mu}^*(\bx,t\in
C).\l{2.16b}\ee

The "closed-path" boundary conditions: \be a_{a\mu}(x\in\s_\infty)
=0,\l{2.16a}\ee where $\s_\infty$ is the remote time-like
hypersurface, is assumed. We will demand that the surface terms
are cancelled in the difference $S_C(A^+)-S^*_C(A^-)$, i.e. \be
\int dx \pa_\mu(A_{a\nu}\pa^\mu A^{a\nu})^+=\int dx
\pa_\mu(A_{a\nu}\pa^\mu A^{a\nu})^-.\l{}\ee Therefore, not only
the trivial pure gauge fields can be considered on $\s_\infty$.

Expanding $S(A\pm a)$ over $a_{a\mu}$, one can write: \be
S(A+a)-S^*(A-a)=U(A,a)+2\Re\int_C dx a_{a\mu}(x)\f{\d S(A)} {\d
A_{a\mu}(x)}.\l{2.17a}\ee This equality will be used as the
definition of the remainder term, $U(A,a)$. With the $\vep$
accuracy, $U(A,a)=O(a^3)$, i.e $U(A,a)$ introduces the
interactions. Notice that \be \f{\d S(A)} {\d A^{a\mu}(x)}=D^{\nu
b }_aF^b_{\mu\nu}\l{2.19a}\ee is the covariant of gauge
transformations. Therefore, the remainder term, $U(A,a)$, is the
gauge invariant if (\r{2.14}) is held.

Inserting (\r{2.15a}) and (\r{2.17a}) into (\r{2.12}), we find:
\be \cn=\int DA\int Da~e^{2i\Re\int_C dx a_a^\mu(x)D^{\nu b
}_aF^b_{\mu\nu}}~e^{iU(A,a)}.\l{2.20a}\ee The integrals over
$a_{a\mu}(x)$ will be calculated perturbatively. For this purpose
one can use the identity: \be
e^{iU(A,a)}=\lim_{\z_{a\mu}=J_{a\mu}=0}e^{-i\kb(J,\z)}e^{2i\Re\int_C
dx a_{a\mu}(x) J^{a\mu}(x)}e^{iU(A,\z)},\l{2.21}\ee where%
\be 2\kb(J,\z)= \Re\int_C dx\f{\d}{\d J^{a\mu}(x)}\f{\d}{\d
\z_{a\mu}(x)}.\l{2.22a}\ee In the future we will omit  the sign of
the limit bearing in mind the prescription: the auxiliary
variables, $J_{a\mu}$ and $\z_{a\mu}$, must be taken equal to zero
at the very end of calculations.

Assuming that the perturbation series will exist, the insertion of
the Eq.(\r{2.21}) into (\r{2.20a}) gives the desired expression:
\be \cn=e^{-i\kb(J,\z)}\int DM(A)e^{iU(A,\z)},\l{2.23}\ee where $$
DM(A)=\prod_{\bx,t\in
C+C^*}\prod_{a,\mu}dA_{a\mu}(x)\int\prod_{\bx,t \in C+C^*}
\prod_{a,\mu}\f{da_{a\mu}(x)} {\pi}e^{2i\Re\int_c dx
a^{a\mu}(D^{\nu b
}_aF^b_{\mu\nu}-J_{a\mu}(x))}=$$\be=\prod_{\bx,t\in
C+C^*}\prod_{a,\mu}dA_{a\mu}(x)\d\le(D^{\nu b
}_aF^b_{\mu\nu}-J_{a\mu}(x)\ri)\l{2.24}\ee is the functional Dirac
measure. The functional $\d$-function on the complex time contour
$C+C^*$ has the definition: \be \prod_{\bx,t\in C+C^*} \d(D^{\nu b
}_aF^b_{\mu\nu}-J_{a\mu})=\prod_{\bx,t\in C}\d(\Re(D^{\nu b
}_aF^b_{\mu\nu}-J_{a\mu})) \d(i\Im(D^{\nu b }_aF^b_{\mu\nu}-
J_{a\mu}))\l{2.26}\ee where the equality (\r{2.16b}) was used.

It can be shown that (\r{2.23}) gives the ordinary perturbation
theory (pQCD) \C{jmp-1,jackiw-1, jackiw-2, faddeev} if the
equation: \be D^\nu_{ab}F^b_{\mu\nu}= J_{a\mu} \l{2.25}\ee is
expanded in the vicinity of $A_{a\mu}=0$. Notice that the
Eq.(\r{2.25}) is not gauge covariant because of $J_{a\mu}(x)$.

\subsection{\it Definition of physical coset space}

The approach based only on the Dirac measure, $DM$, is incomplete
since all strict solutions $u^i_{a\mu}$ of the equation (\r{2.25})
must be taken into consideration: \be \cn=\sum_i \cn_i,\l{2.20}\ee
where $\cn_i$ corresponds to $u^i_{a\mu}$. It is necessary to
extract one, physical, term in (\r{2.20}) since each $i$-th
solution belongs to the different symmetry class \C{smale}.

Notice that the non-diagonal terms, \be Z_iZ_j^*+Z_i^*Z_j,~ i\neq
j,\l{2.27}\ee are absent in the sum (\r{2.20}) as the consequence
of orthogonality of the Hilbert spaces, see (\r{2.24}). This
allows to offer the following selection rule.

\vskip 0.25cm {\bf Corollary. \it If there are no special external
conditions and if the coset space $W_i$ is adjusted to
$u^i_{a\mu}$ then the dimension of the physical coset space, $\dim
W$, is defined by the condition: \be\dim W=\max\{\dim W_i\}.
\l{2.19}\ee The solution of (\r{2.25}) must be chosen in
accordance with this selection rule.}

Indeed, having (\r{2.20}) and noting the absence of the
non-diagonal terms of (\r{2.27}) type, one may use the notion of
the "situation of general position" ordinary for classical
mechanics \C{arnold, abraham}. It means the absence of special
boundary conditions to the Eq. (\r{2.25}). Then in the sum over
the strict solutions of the Lagrange equation, namely, the one
which gives the largest contribution, $u_{a\mu}$, must be left.
Other contributions would be realized on the zero measure since
the (\r{2.20}) includes summation over initial conditions\foot
{For instance, the trivial trajectory corresponds to the particle
lying motionless in the semiclassical approximation at the bottom
of a potential hole. This type of trajectories belongs to a
definite topology class \C{smale} and must be refused as it
follows from our selection rule, Corollary.}, see Sec.4.3.

\vskip0.25cm Following the selection rule (\r{2.19}) the Gribov
ambiguity actually presents the problem in the non-Abelian gauge
theory since we know, at least, the $O(4)\t O(2)$-invariant strict
solution of the $SU(2)$ Yang-Mills equation. The corresponding
coset space has $\dim W=8$ plus the (infinite) gauge groups
dimension. Therefore, following the selection rule (\r{2.19}), the
excitations in the vicinity of $A_{a\mu}=0$ are realized on the
zero measure.

The $ansatz$ \C{t'hooft-1, uy}: \be \sqrt{g}u_\mu^a= \q^a_{\mu\nu}
\pa^\nu\ln u\l{2.2}\ee  for the $SU(2)$ Yang-Mills potential
$u_\mu^a$ leads to the conformal scalar field theory, see e.g.
\C{actorr}. For our purpose it is enough to know the existence of
the exact solution \C{deAlfaro, actorr}: \be
u(\bx,t)=\le\{\f{(\ga- \ga^*)^2}
{(x-\ga)^2(x-\ga^*)^2}\ri\}^{1/2},\l{2.3a}\ee where \be
\ga_\mu=\x_\mu+i\q_\mu,~~\ga_\mu\ga^\mu=g^{\mu\nu}\ga_\mu\ga_\nu=
\ga_0^2-\ga_i^2,~i=1,2,3\l{}\ee and $\x_\mu$ and $\q_\mu$ are the
real numbers. Their physical domain is defined by inequalities:
\be -\infty\leq\x_ \mu \leq+ \infty,~~-\infty\leq\ga_i \leq+
\infty,~~\sqrt{\q_\mu\q^\mu}= \q\geq0. \l{}\ee The latter
condition means that the energy of $u_{a\mu}$ is nonnegative.

The solution (\r{2.3a}) is regular in the Minkowski metric for
$\q>0$ and has the light-cone singularity at $\q=0$, i.e the
solution is singular at the border $\pa_\q W$. We will regularize
it continuing contributions on the Mills complex-time contour,
Sec.2.1. The solution (\r{2.3a}) has the finite energy and no
topological charge. There also exist its elliptic generalizations
\C{actorr}

The selection rule (\r{2.19}) defines the ground state of the
theory on the Dirac measure. Thus, a chosen vacuum optionally has
the lowest energy. This definition of the ground state is useful
if the finite-time dynamic problem is considered.

\section{Mapping into the coset space}\0

Considering the general transformation: \be (A_{a\mu},P_{a\mu})
(\bx,t)\to (\la,\k)_\a(t),\l{3.3a}\ee  where $A_{a\mu}$ is the
arbitrary set of fields and $P_{a\mu}$ is the conjugate momentum,
one must conserve the dimension of the path integral measure: \be
\dim DM(A_{a\mu},\pi)=\dim DM(\la,\k),\l{3.2c}\ee where $$
DM(\la,\k)=\prod_{\a,t}d\la_\a(t) d\k_\a(t).$$ Therefore, the
Eq.(\r{3.2c}) defines the set $\{\a\}$. For the Yang-Mills theory
the set $(\la,\k)_\a$ includes the gauge phase $\La_a(\bx,t)$. For
a more confidence, one may consider the theory on the space
lattice.

It is assumed that the time dependent variables will be defined on
the complete Mills time contour, $C+C^*$. We will formulate the
general method of mapping (\r{3.3a}) into the infinite dimensional
phase space $\Ga_\infty$, Sec.3.2, and then will find the
reduction procedure, $\Ga_\infty\mapsto W$ on the second stage of
the calculation, Sec.4.1.

\subsection{\it First order formalism}

The action in terms of the electric field, $E^i_a=F^{i0}_a,~
i=1,2,3,$ looks as follows:\be S(\ma,\me)=\int dx
\le\{\dot{\ma}_a\cd\me_a + \f{1}{2}(\me_a^2+\mathbf{B}_a^2)-
A_{0a}(\cd\me)_a\ri\},\l{p2}\ee where the magnetic field $
\mathbf{B}_{a}(\ma)= {rot}\ma_{a}+ \f{1}{2}(\ma\t \ma)_a$. The
corresponding Dirac measure is: $$ DM(\ma,\mp)=\prod_{a}\prod_{x}
d\ma_{a}(x)~d\mp_{a}(x)~\d(\cd\mp_b)\t $$\be \t\d\le(\dot\mp_a(x)+
\f{\d H_J(\ma,\mp)}{\d\ma_a(x)}\ri) \d\le(\dot\ma_a(x)-\f{\d
H_J(\ma,\mp)}{\d\mp_a(x)}\ri), \l{f10a}\ee where
$d{\mathbf{A}}_a(x)d{\mathbf{P}}_{a}(x)=\prod_idA_{ia}
d{{P}}_{ai}(x),~i=1,2,3,$  and the total Hamiltonian  \be%
H_J=\f{1}{2}\int d^3x({\mathbf P}^2_a+{\mathbf B}_a^2) +\int d^3x
{\mathbf j}_a {\mathbf A}_a.\l{a4}\ee Notice that the dependence
on $A_{a0}$ was integrated out and the Gauss law, $\cd^a_b{\mathbf
P}_b=0,$ was appeared as a result in (\r{f10a}). The Faddeev-Popov
$ansatz$ was not used for the definition of the integral over
$A_{a\mu}$. The perturbations generating operator $\kb$ and the
remainder potential term $U$ stay unchanged, see (\r{2.22a}) and
(\r{2.17a}).

The integrals with the measure (\r{f10a}) will be calculated using
new variables.

\subsection{\it General mechanism of transformations}

{\bf Proposition 1. } {\it The Jacobian of transformation of the
Dirac measure is equal to one} \C{jmp-1}.\\ One can insert the
unite $$1=\f{1}{\D(\la,\k)}\int\prod_{\a,t}d\la_\a(t)d\k_\a(t)\t
$$\be \t\prod_{a,\bx}\d(\ma_a(\bx,t)- \bu_a(\bx;\la(t),\k(t)))~
\d(\mp_a(\bx,t)-\bp_{a}(\bx;\la(t),\k(t))), \l{3.2}\ee into the
integral (\r{2.23}) and integrate over $\ma_a$ and $\mp_a$ using
the $\d$-functions of (\r{3.2}). In this case the transformation
is performed. Otherwise, if the $\d$-functions of (\r{f10a}) are
used, $\bu_a$ and $\bp_a$ will play the role of constraints and
(\r{3.2}) will present the Faddeev-Popov $ansatz$. It must be
noted that the both ways of calculation must lead to the identical
ultimate result because of the $\d$-likeness of measures in
(\r{f10a}) and (\r{3.2}). The first way is preferable since it
does not imply the ambiguous gauge fixing procedure \C{gribov,
atiyah, zinger}.

The arbitrary given composite functions
$\bu_{a}(\bx;\la(t),\k(t))$ and $\bp_{a}(\bx; \la(t),\k(t))$ must
obey the condition:\be \D(\la,\k)=\int \prod_{\a,t} d\la'_\a(t)
d\k'_\a(t)\prod_{a,\bx} \d(\la'\bu_{a,\la}+\k'\bu_{a,\k})
\d(\la'\bp_{a,\la} +\k'\bp_{a,\k}) \neq0,\l{3.5}\ee where $$
\bu_{a,X}\equiv\f{\pa \bu_{a}}{\pa X}, ~\bp_{a,X}\equiv\f{\pa
\bp_{a}}{\pa X},~~X=(\la_\a,\k_\a).$$ The summation over the
repeated index, $\a$, will be assumed.

The transformed measure: $$ DM(\la,\k)=\f{1}{\D(\la,\k)}
\prod_{\a,t} d\la_\a(t)d\k_\a(t)\t$$
$$\t\prod_{a,\bx}\d\le(\dot{\la}\bu_{a,\la}+ \dot{\k}
\bu_{a,\k}-\f{\d H_J(\bu,\bp)}{\d
\bp_{a}}\ri)d\le(\dot{\la}\bp_{a,\la}+ \dot{\k}\bp_{a,\k}+\f{\d
H_J(\bu,\bp)}{\d \bu_{a}}\ri)$$ can be diagonalized introducing
the auxiliary function(al) $h_J$: $$ DM(\la,\k)=
\f{1}{\D(\la,\k)}\prod_{\a,t} d\la_\a(t)d\k_\a(t) \int\prod_{\a,t}
d\la_\a'(t) d\k_\a'(t)\t $$ $$\t
\d\le(\la_{\a}'-\le(\dot{\la}_\a-\frac{\d h_J(\la,\k)}{\d
\k_\a}\ri)\ri) \d\le(\k_\a'-\le(\dot{\k}_\a+\f{\d
h_J(\la,\k)}{\d\la_\a}\ri)\ri)\t $$\be\t\prod_{a,\bx}
\d\le(\bu_{a,\la}\la'+\bu_{a,\k}\k'+\{\bu,h_J\}_a -\f{\d H_J}{\d
\bp_a}\ri)\d\le(p_{a,\la} \la'+p_{a,\k}\k'+\{\bp,h_J\}_a+\f{\d
H_J}{\d \bu_a}\ri),\l{3.6}\ee where $\{,\}$ is the Poisson
bracket.

Let us assume now that $\bu_a$, $\bp_a$ and $h_J$ are chosen in
such a way that: \be \{\bu_a,h_J\}-\f{\d H_J}{\d
\rp_a}=0,~~\{\bp_a,h_J\}+ \f{\d H_J}{\d \ru_a}=0.\l{3.7}\ee Then,
having the condition (\r{3.5}), the transformed measure takes the
form, see (\r{3.6}): \be DM(\la,\k)= \prod_{\a,t} d\la_\a(t)
d\k_\a(t) \d\le(\dot{\la}_\a-\frac{\d h_J(\la,\k)}{\d \k_\a}\ri)
\d\le(\dot{\k}_\a+\f{\d h_J(\la,\k)}{\d\la_\a}\ri), \l{3.8}\ee
where the functional determinant $\D(\la,\k)$ was cancelled.

As a result, \be \cn=e^{-i\kb(J,\z)}\int DM(\la,\k) e^{i
U(\bu,\z)} \l{3.15a}\ee where $\kb(J,\z)$ was defined in
(\r{2.22a}), $DM(\la,\k)$ was defined in (\r{3.8}) and $U(\bu,\z)$
was introduced in (\r{2.17a}). Therefore, the Jacobian of
transformation is equal to one, i.e. in the frame of the
conditions (\r{3.5}) and (\r{3.9}) the phase space volume is
conserved. Q.E.D. \vskip0.3cm

We will consider the case when $h_J$ is the linear over $J(\bx,t)$
functional: \be h_J(\la,\k)=h(\la,\k)+\int d\bx \mj_a(\bx,t)
\my_a(\bx;\la,\k),\l{3.15}\ee where $\my_a$ are the arbitrary vector
functions. Transforming the theory we get to the dynamical problem
for $\la_\a(t)$ and $\k_\a(t)$: $$\dot{\la}_\a=\frac{\d
h_J(\la,\k)}{\d \k_\a}=\frac{\d h(\la,\k)}{\d \k_\a}+\int d\bx
\mj_a\f{\d\my_a}{\d \k_\a}\equiv h_{\k_\a}+\int d\bx \mj_a\my_{a,\k}
,$$ \be\dot{\k}_\a=-\f{\d h_J(\la,\k)}{\d\la_\a}= -\f{\d
h(\la,\k)}{\d\la_\a}-\int d\bx \mj_a\f{\d\my_a}{\d\la_\a}\equiv
-h_{\la_\a}-\int d\bx \mj_a\my_{a,\la}. \l{3.13}\ee The equality
(\r{3.15}) was used here.

\vskip0.25cm{\bf Proposition 2. \it If (\r{3.15}) is held then the
transformation {\rm(\r{3.3a})} induces the splitting: \be \mj_a\to
\{j_\la,j_\k\}\l{3.17}\ee} The proof of the splitting comes from
the identity: $$\prod_{\a,t}\d\le(\dot{\la}_\a-\frac{\d
h_J(\la,\k)}{\d \k_\a}\ri) \d\le(\dot{\k}_\a+\f{\d
h_J(\la,\k)}{\d\la_\a}\ri)=$$ $$=
\exp\{-i\bk(j,e)\}\exp\le\{2i\Re\int_C d\bx dt\mj_a(\bx,t)
(e_\la\my_{a,\k}+e_\k\my_{a,\la})\ri\}\t$$ \be\t \prod_{\a,t}
\d(\dot{\la}_\a-h_{\k_\a}-j_{\la_\a}) \d(\dot{\k}_\a+
h_{\la_\a}+j_{\k_\a}),\l{3.14}\ee where \be 2\bk(j,e)=\Re\int_C dt
\le(\f{\d}{\d j_{\la}}\f{\d}{\d e_{\la}} +\f{\d}{\d
j_{\k}}\f{\d}{\d e_{\k}}\ri).\l{3.15b}\ee At the very end one must
take $j_X=e_X=0$, $X=(\la,\k)$. The equality (\r{3.14}) can be
derived using the functional $\d$-functions Fourier transformation
(\r{2.24}).

Inserting (\r{3.14}) into (\r{3.15a}) we find the completely
transformed representation for $\cn$, where the individual to each
degree of freedom quantum sources, $j_X$, $X=(\la,\k)$, appears.
The transformed representation of $\cn$ looks like:
\be\cn=e^{-i\bk(j,e)} \int DM(\la,\k)e^{iU(\bu,\bE)}, \l{3.20}\ee
where \be DM(\la,\k)= \prod_{\a, t}d\la_\a(t) d\k_\a(t)
\d(\dot\la-h_{\k}(\la,k)-j_\la) \d(\dot\k+
h_\la(\la,k)+j_\k),\l{3.18}\ee \be \bE_a=
e_{\la}\my_{a,\k}+e_{\k}\my_{a,\la}\l{3.22}\ee and $\bk(j,e)$ was
defined in (\r{3.15b}).

\vskip0.25cm{\bf Proposition 3. \it The Eq.{\rm(\r{3.7})} and the
measure {\rm(\r{3.18})} define the classical flow for arbitrary
$h_J(\la,\k)$.}\\ Indeed, $$ \dot{\bu}_a= \dot{\la}\bu_{a,\la}+
\dot{\k}\bu_{a,\k}=\{\bu_a,h_J\}=\f{\d H_J}{\d\mp_a}, $$\be
\dot{\bp}_a=
\dot{\la}\bp_{a,\la}+\dot{\k}\bp_{a\,k}=\{\bp_a,h_J\}=- \f{\d
H_J}{\d \bu_a}, \l{3.12}\ee where (\r{3.18}) and then (\r{3.7})
have been used step by step. Therefore, $\bu_{a}$ is the solution
of the sourceless Lagrange equation (\r{2.25}) and $\bp_a=
\dot{\bu}_a$. \vskip0.25cm

We will consider the following solution of (\r{3.7}): \be
h_J(\la,\k)=H_J(\ru,\rp), \l{3.9}\ee i.e. the case where $h$ is
the transformed Hamiltonian and \be
\my_a(\bx;\la,\k)=\bu_a(\bx;\la,\k).\l{3.24}\ee

\section{Reduction}\0

\subsection{\it Cyclic variables}

Proposition 3 means that $\a$ in (\r{3.3a}) is the coset space
index. Let us divide the set $\{\la,\k\}$ into two parts:
\be\{\la,\k\}\to(\{\la,\k\},\{\x',\q'\}),\l{3.19a}\ee assuming
that $\la$ and $\k$ are cyclic variables: \be
\f{\pa\bu_a}{\pa\la}=
O(\e),~~\f{\pa\bu_a}{\pa\k}=O(\e),~\e\to0,\l{3.21}\ee and the
derivatives of $\bu_a$ over $\x'$ and $\q'$ are finite at $\e=0$.
It can be shown that the variables $(\la,\k)$ stay cyclic in the
quantum sense as well.

\vskip0.25cm {\bf Proposition 4. \it The quantum force is
orthogonal to the cyclic variables axes.}\\ Indeed, taking into
account (\r{3.21}), \be \bk(j,e)=\int dt \le\{\f{\d}{\d
j_{\la}}\cdot \f{\d}{\d e_{\la}}+ \f{\d}{\d j_{\k}}\cdot \f{\d}{\d
e_{\k}}+\f{\d}{\d j_{\x'}}\cdot \f{\d}{\d e_{\x'}}+ \f{\d}{\d
j_{\q'}}\cdot \f{\d}{\d e_{\q'}}\ri\}.\l{4.3}\ee As it follows
from (\r{3.18}), \be \f{\d \bu_a}{\d j_X}\sim\f{\d\bu_a}{\d
X}=O(\vep),~X=(\la,\k).\l{}\ee Therefore, we can write in the
limit $\vep=0$ that \be 2\bk(j,e)=\int dt \le\{\f{\d}{\d
j_{\x'}}\cdot \f{\d}{\d e_{\x'}}+ \f{\d}{\d j_{\q'}}\cdot
\f{\d}{\d e_{\q'}}\ri\}.\l{4.5}\ee Then, following our definition,
one should take everywhere \be j_X=e_X=0,~X=(\la,\k).\l{}\ee
Q.E.D.

\vskip 0.25cm The result of the reduction looks as follows:\be
DM(u,p)=d\O~ DM(\x',\q'),\l{4.7}\ee where the infinite dimensional
integral over \be d\O=\prod_{\a,t} d\la_\a(t)
d\k_\a(t)\d(\dot\la_\a(t)) \d(\dot\k_\a(t))\l{4.8}\ee will be
cancelled by normalization. This procedure completes the
renormalization of the transformed formalism.

The remaining degrees of freedom are entered into the reduced
Dirac measure: \be DM(\x',\q')=\prod_t d\x'(t)d\q'(t)
\d(\dot\x'-h_{\q'}(\x',\q')-j_{\x'}) \d(\dot\q'+
h_{\x'}(\x',\q')+j_{\q'}).\l{7a}\ee This result presents the first
step of the reduction into the physical coset space $W$.

\subsection{\it Quantization rule in the coset space}

The case when only the part of variables are cyclic:
$\{\x'\}=(\{\x\},\{\x''\})$ and $\{\x'\}=(\{\q\},\{\q''\})$, \be
\dim\{\x\}=\dim\{\q\},\l{4.9a}\ee where only $\{\x''\}$ is the set of
cyclic variables: \be \f{\pa\bu_a}{\pa\x''}= \e, \l{4.9}\ee comes
within the conditions of Proposition 4.

Then we can define the set $\{\q\}$ under the condition: $\pa
h/\pa\q\neq0$. In the frame of this definition $\q_\a$ are the
integrals of motion. This gives: \be DM(\x,\q'',\q)=\prod_t
d\q''(t)d\q'(t)d\x(t)
\d(\dot\q''-j_{\q''})\d(\dot\q+j_{\q})\d(\dot\x-h_{\q}(\q)-j_{\x}).
\l{4.10}\ee

Following (\r{3.22}) and (\r{3.24}) the virtual deviation $\bE$
looks as follows: \be \bE_a= e_{\x}\bu_{a\q}+e_{\q}\bu_{a\x}
+e_{\q''}\bu_{a\x''} \l{}\ee and the perturbations generating
operator is: \be 2\bk(j,e)=\int dt \le\{\f{\d}{\d j_{\x}}
\f{\d}{\d e_{\x}}+ \f{\d}{\d j_{\q}}\f{\d}{\d e_{\q}}+ \f{\d}{\d
j_{\q''}} \f{\d}{\d e_{\q''}}\ri\}.\l{4.12a}\ee As it follows from
the general condition that the auxiliary variables must be taken
equal to zero, we must put $e_{\q''}=0$ since (\r{4.9}). We must
omit simultaneously the last term in (\r{4.12a}). For this reason
one must put $j_{\q''}=0$ in (\r{4.10}).

\vskip0.25cm {\bf Proposition 5. \it The coset space quantized
variables form the even dimensional symplectic manifold,
$\{\x,\q\}\in T^*W$, with the canonical equal-time Poisson
brackets:\be \{\bu(\bx;\x,\q), \bu(\by;\x,\q)\}=
\{\bp(\bx;\x,\q),\bp(\by;\x,\q)\}=0,~~
\{\bu(\bx;\x,\q),\bp(\by;\x,\q)\}=\d_{\bx,\by} \l{4.14}\ee iff
$\{\bx\}\nsubseteq\{\a\}$.} One must insert (\r{3.9}) into
(\r{3.7}) in order to prove this proposition. \vskip0.25cm

Proposition 5 means that $\{\La_a\}\nsubseteq\{\x,\q\}$.

\subsection{\it Concluding expression}

As a result, \be \cn=e^{-i\mathbf{k}(je)} \int DM(\x,\q)
e^{iU(\bu,\bE)}, \l{18}\ee where the new coset space virtual
deviation is \be \bE_{a}=\sum_\a\le\{e_{\x_\a}\f{\pa \bu_{a}}
{\pa\q_\a}+e_{\q_\a}\f{\pa \bu_{a}}{\pa\x_\a}\ri\}.\l{20}\ee

The generating quantum perturbations operator in the coset space
is \be 2\mathbf{k}(je)=\sum_\a\int dt\le\{\f{\d}{\d
j_{\x_\a}(t)}\f{\d}{\d e_{\x_\a}(t)}+\f{\d}{\d j_{\q_\a}(t)}
\f{\d}{\d e_{\q_\a}(t)}\ri\}, \l{4.11}\ee where summation is
performed over all canonical pairs, $(\x,\q)\in T^*W$. The
corresponding measure \be DM=dR\prod_{\a,t}
d\x_\a(t)d\q_\a(t)\d\le(\dot\x_\a- h_{\q_\a}(\q)-j_{\x_\a}\ri)
\d\le(\dot\q_\a+j_{\q_\a}\ri), \l{4.12}\ee where $dR$ is the zero
modes Cauchy measure: \be dR=\prod_{\a,t} d\q_\a''\d(\dot\q_\a'').
\l{4.13}\ee Therefore, \be W=T^*W+R\l{}\ee where $\{\x,\q\}\in
T^*W$ and $\{\x''\}\in R$.

The coset space Hamiltonian equations: \be \dot\x_\a-
h_{\q_\a}(\q)=j_{\x_\a},~\dot\q_\a=-j_{\q_\a}\l{4.22}\ee are
easily solved through the Green function $g(t-t')$. The latter
must obey the equation: \be \pa_tg(t-t')=\d(t-t').\l{4.23}\ee This
Green function has the universal meaning, and it must be the same
for the arbitrary theory. Then, using the $i\vep$-prescription and
the experience of the Coulomb problem considered in \C{jmp-1}, we
will use the following solution of (\r{4.23}):\be g(t)=\le\{
\begin{array}{c}
  1,~~ t\geq0 \\
  0,~~ t<0
\end{array}.\ri.\l{}\ee The solution of the Eq.(\r{4.22}) looks
as follows: \be \x_a^j(t)=\int dt' g(t-t')\{h_{\q_\a}(\q^j)+
j_{\x_\a}\}(t'),~~\q_a^j(t)=-\int dt g(t-t')j_{\q_\a}(t'). \l{}\ee
As a result, the functional measure $DM$ is reduced to the Cauchy
measure \be dM=\prod_{\a,t} d\q_\a''(t)\d(\dot\q_\a'')
d\x_\a(t)d\q_\a(t)\d(\dot\x_\a) \d(\dot\q_\a)= \prod_{\a}
d\q_\a''(0)d\x_\a(0)d\q_\a(0).\l{}\ee The integral over $dM$ gives
the volume ${V}$ of the factor group $\cg/\ch$ and $\ln{V}\leq\dim
W$. Notice that the gauge group volume ${V}_\La$ in our formalism
is defined by the measure $\prod_{a,\bx} d\La_a(\bx,0)$.

Therefore, \be \cn=e^{-i\mathbf{k}(je)} \int dM
e^{iU(\bu^j,\bE^j)}, \l{4.27}\ee where $\bu^j$ and $\bE^j$ are
dependents on $(\x_a^j, \q_a^j)$ functions.

\subsection{\it Gauge invariance}

Following (\r{2.17a}) and (\r{20}), $$U(u,e)=S(u+e)-S^*(u-e)-2\Re
\sum_{a,\a} \int_C dx \le\{e_{\x_\a}\f{\pa \ru_{ai}(x)}{\pa\q_\a}+
e_{\q_\a} \f{\pa \ru_{ai}(x)}{\pa\x_\a}\ri\}\f{\d S(u)}{\d
u_{ai}(x)} =$$\be=S(u+e)-S^*(u-e)-2\Re\sum_{\a}\int_C dt
\le\{e_{\x_\a}(t)
\f{\d}{\d\q_\a(t)}+e_{\q_\a}(t)\f{\d}{\d\x_\a(t)}
\ri\}S(u).\l{}\ee This quantity is transparently gauge invariant.

We can conclude that each term of the coset space perturbation theory
is gauge invariant since $DM$ in (\r{4.12}) and $\bk(je)$ in
(\r{4.11}) are the gauge invariant quantities.

It is interesting to note that in spite of the fact that each term
of the perturbation theory is transparently gauge invariant,
nevertheless, one can not formulate the theory in terms of the
gauge field strength.

\section{Conclusions}

It is useful to summarize the rules of the coset space perturbation
theory.

(i) The transformation, (\r{3.3a}), to independent variables,
(\r{3.5}), is performed having in mind that the power of the
variables set should not be altered, see (\r{3.2c}).

(ii) The "host free" transformation is induced by the function
$\bu_a$ defined by the Eq.(\r{3.7}). In this stage the function
$h_j(\la,\k)$ is arbitrary.

(iii) If $h_J$ is chosen as the liner function of $\mj_a$,
$h_J=h+O(J)$ then there exists a mapping into the $(\la,\k)$
space, see (\r{3.17}). This mapping produces a new set of sources
$\{j_\la,j_\k\}$, (\r{3.15b}), and virtual deviations,
$\{e_\la,e_\k\}$, (\r{3.22}). It is remarkable that each degree of
freedom of the $(\la,\k)$ space is excited independently of one
another by the individual sources $\{j_\la,j_\k\}$. This is
crucial for the reduction of the quantum degrees of freedom.

(iv) One can consider the case when a subset of variables is
cyclic, (\r{3.19a}), (\r{3.21}). The latter means that if
$h=h+O(\e)$, $\e\to0$, then because of the perturbation term,
$O(\e)$, $\bu_a$ occupies the whole infinite dimensional space
$\Ga_\infty$. Within  the limit of $\e=0$ the trajectory subsides
on the surface, $W$, of a smaller dimension. The redundant
variables at $\e=0$ are cyclic. As a result we have found the
reduced measure (\r{7a}), and the perturbations generating
operator (\r{4.5}). The volume of the cyclic variables, (\r{4.8}),
is cancelled by normalization. The field theoretical problem
becomes finite dimensional. The cancellation of the cyclic
variables volume can be considered as a renormalization procedure.

(v) One may consider now the case when $h_J$ is the transformed
Hamiltonian. For this reason $W$ is the coset space. The physical
value of $\dim W$ is defined by Corollary. In other respects the
choice of the coset variables $\{\x,\q\}$ is arbitrary.

(vi) A portion of the remaining variables can belong to the
symplectic subspace $T^*W\subseteq W$, with the Poisson brackets
(\r{4.14}). The latter allows to conclude that the gauge phase
$\La_a$ can not belong to $T^*W$. As a result the perturbation
theory is transparently gauge invariant.

(vii) The known solution \C{deAlfaro} shows that all space-time
integrals of the coset space perturbation theory are finite
outside the border $\pa W$ since $|S(u)|<\infty$ and $\dim W$ is
finite. The border contributions, $\sup(\x,\q)\in \pa W$, remain
finite because of the $i\vep$-prescription. Further analysis of
the role of the border singularities, see also \C{jmp-1}, will be
given in subsequent publications.

\vskip 0.4cm {\large \bf Acknowledgement}

We would like to thank V. Kadyshevsky and our colleagues in the
Lab. of Theor. Phys. of JINR for their deep interest in the
described approach.


\vspace{0.25in}
\begin{thebibliography}{99}

\bibitem{abraham} Abraham, R., Marsden, J.E.: {\it Foundations of
Mechanics}. Benjamin/Cummings Publ. Comp., Reading, Mass., 1978

\bibitem{actorr} Actor, A.: Classical solutions of SU(2) Yang-Mills
theories. Rev. Mod. Phys. {\bf51}, 461-525 (1979)

\bibitem{deAlfaro} DeAlfaro, V., Fubini, S. Furlan, G.: A new
classical solution of the Yang-Mills field equations. Phys. Lett.
{\bf B65}, 163-166  (1976)

\bibitem{arnold} Arnold, V.I.: {\it Mathematical Methods of
Classical Mechanics}. Springer Verlag, New York, 1978

\bibitem{atiyah} Atiyah, M.F., Jones, J.D.S.: Topology aspects
of Yang-Mills theory. Commun. Math. Phys. {\bf61}, 97-118 (1978)

\bibitem{dirak} Dirac, P.A.M.: {\it Lectures on Quantum
Mechanics}. Yeshiva Univ., New York, 1964

\bibitem{edwards} Edwards, S.F., Gulyaev, Y.V.: Path integrals
in polar co-ordinates. Proc. Roy. Soc. {\bf A279}, 229-235 (1964)

\bibitem{faddeev}  Faddeev, L.D., Korepin, V.E.: Quantum theory of
solitons, preliminary version. Phys. Rept. {\bf42}, 1-87 (1978)

\bibitem{fedor} Fedoryuk, M. V.: {\it Asymptotic: integrals and
series}. M: Nauka, 1987

\bibitem{gribov} Gribov, V.N.: Quantization of non-Abelian gauge
theories. Nucl. Phys. {\bf B139}, 1-19 (1978)

\bibitem{hioe} Hioe, F.T., Macmillen, D., Montroll, E.W.: Quantum
theory of anharmonic oscillators. Phys. Rept. {\bf 43}, 305-335
(1978)

\bibitem{t'hooft-1} 't Hooft, G.: Computation of the quantum
effects due to a four-dimensional pseudoparticle. Phys. Rev. {\bf D
14}, 3432-3450 (1976)

\bibitem{t'hooft-2} 't Hooft, G.: Borel summability of a
four-dimensional field theory. Phys. Lett. {\bf B119}, 369  (1982)

\bibitem{isham} Isham, C.J.: {\it Relativity, Groups and Topology}
II. Eds. B.S. De Witt and R. Stora, North-Holland, Amsterdam, 1984

\bibitem{jackiw-1} Jackiw, R.: The quantum theory of solitons and
other nonlinear classical waves. Phys. Rept. {\bf23}, 273-280
(1976)

\bibitem{jackiw-2} Jackiw, R.: Quantum meaning of classical field
theory. Rev. Mod. Phys. {\bf 49}, 681-706 (1977)

\bibitem{linden} Landsman, N.P., Linden, N.: Geometry of inequivalent
quantization. Nucl. Phys. {\bf B365}, 121-160 (1991)

\bibitem{landsman} Landsman, N.P.: Quantization and
superselection sectors. Rev. Math. Phys. {\bf 2}, 45-72, 73-104
(1991)

\bibitem{mackey}Mackey, G.W.: {\it Induced Representation of Groups
and Quantum Mechanics}. Benjamin, New York, 1969

\bibitem{yaph} Manjavidze, J.: About connection among classical and
quantum descriptions. Sov. J. Nucl. Phys. {\bf45}, 442-451 (1987)

\bibitem{elpat} Manjavidze, J.: Wigner functions of essentially
nonequilibrium systems. Phys. Part. Nucl. {\bf 30}, 49-65 (1999),
hep-ph/9802318

\bibitem{jmp-1} Manjavidze, J.: Topology and perturbation theory.
J. Math. Phys. {\bf41}, 5710-6734 (2000)

\bibitem{jmp-3} Manjavidze, J., Sissakian, A.: Yang-Mills fields
quantization in the factor space. J. Math. Phys. {\bf42},
4158-4180 (2001)

\bibitem{physrep} Manjavidze, J., Sissakian, A.: Very high
multiplicity processes. Phys. Rept. {\bf346}, 1-88 (2001)

\bibitem{tmf} Manjavidze, J., Sissakian, A.: A Field theory
description of constrained energy-dissipation processes. Theor. Math.
Phys. {\bf 130}, 153-197 (2002)

\bibitem{marinov} Marinov, M.S.: Path integrals in quantum theory:
an outlook of basic concepts. Phys. Rep. {\bf 60}, 1-57 (1980)

\bibitem{milss} Mills, R.: {\it Propagators for Many-Particle
Systems}. Gordon {\&} Breach, Science, NY, 1970

\bibitem{zinger} Singer, I.M.: Some remarks on the Gribov ambiguity.
Comm. Math. Phys. {\bf 60}, 7-12 (1978)

\bibitem{shabanov} Shabanov, S.V.: Geometry of the physical phase
space in quantum gauge systems. Phys. Rep. {\bf 326}, 1-163 (2000)

\bibitem{smale} Smale, S.: Inv. Math., {\bf 11:1}, 45 (1970)

\bibitem{uy} Uy, Z.E.S.: A solution to the SU(2) classical
sourceless gauge field equation. Nucl. Phys. {\bf B11}, 389-396
(1976)

\end{thebibliography}
\end{document}